# A Supremum-Norm Based Test for the Equality of Several Covariance Functions

Jia Guo, Bu Zhou, and Jin-Ting Zhang

National University of Singapore

## Abstract

In this paper, we propose a new test for the equality of several covariance functions for functional data. Its test statistic is taken as the supremum value of the sum of the squared differences between the estimated individual covariance functions and the pooled sample covariance function, hoping to obtain a more powerful test than some existing tests for the same testing problem. The asymptotic random expression of this test statistic under the null hypothesis is obtained. To approximate the null distribution of the proposed test statistic, we describe a parametric bootstrap method and a non-parametric bootstrap method. The asymptotic random expression of the proposed test is also studied under a local alternative and it is shown that the proposed test is root-$n$ consistent. Intensive simulation studies are conducted to demonstrate the finite sample performance of the proposed test and it turns out that the proposed test is indeed more powerful than some existing tests when functional data are highly correlated. The proposed test is illustrated with three real data examples.

Short Title: Testing the equality of several covariance functions

## 1 Introduction

Functional data arise in a wide scope of scientific fields such as biology, medicine, ergonomics among others. They are collected in a form of curves or images. Nowadays, this type of data is

First Edition: May 3, 2014, Last Update: September 12, 2016.

Jia Guo (E-mail: jia.guo@u.nus.edu) and Bu Zhou (E-mail: bu.zhou@u.nus.edu) are PhD candidates, Jin-Ting Zhang (E-mail: stazjt@nus.edu.sg) is Associate Professor, Department of Statistics and Applied Probability, National University of Singapore, Singapore 117546. The work was financially supported by the National University of Singapore Academic Research grant R-155-000-164-112.

widely used as basis elements in data analysis instead of traditional scalar or vector observations. Compared with traditional data, functional data are worth exploring since they contain much more structural information. A lot of useful and effective tools have been developed over the past two decades to extract those information hidden in functional data. The reader is referred to Ramsay and Silverman (2005) and Zhang (2013) and references therein for more details.

In the recent decade, much attention about hypothesis testing of the mean functions for functional data has been paid. A general and direct testing procedure is the so-called pointwise $t$-test described by Ramsay and Silverman (2005). However, this pointwise $t$-test cannot give a global conclusion which is often needed in real data analysis. To overcome this drawback, Zhang et al. (2010) proposed an $L^2$-norm based test whose test statistic is obtained as a squared $L^2$-norm of the mean function differences of the two functional samples. For one-way ANOVA problems for functional data, several interesting tests have been proposed in the literature. For example, Zhang and Liang (2013) proposed a GPF test obtained via "globalizing a pointwise $F$-test for functional one-way ANOVA" and Zhang et al. (2016) proposed a test whose test statistic is the supremum value of the pointwise F-test mentioned above, aiming to obtain a more powerful test than the GPF test for one-way ANOVA for functional data. When conducting the above mean function testing, one may first need to check the equality of the covariance functions of the functional samples involved, as many methods for mean function testing are based on the assumption that the functional samples involved have a common covariance function. Motivated by this need of testing the equality of the covariance functions of functional samples, Zhang and Sun (2010), Panaretos et al. (2010) and Fremdt et al. (2013) proposed several useful approaches to address this issue.

Although some recent works have shed some light on how to test the equality of the covariance functions of two functional samples, how to test the equality of the covariance functions of several functional samples receives relatively little attention. Guo et al. (2016) studied some $L^2$-norm based tests. However, these $L^2$-norm based tests may be less powerful when the functional data are highly correlated. In addition, two of the $L^2$-norm based tests developed there are based on the assumption that the functional data are Gaussian. Therefore, they are not applicable when the functional data are non-Gaussian. In this paper, we develop a supremum-norm based test which is good at detecting the covariance function differences when the functional data are highly or moderately correlated and they can also be used for non-Gaussian functional data.

The asymptotic random expression of the proposed supremum-norm based test is derived and its asymptotic power under a local alternative is studied. A parametric and a nonparametric bootstrap tests are described to approximate the associated null distribution. The proposed test is demonstrated via some simulation studies and several real data applications.

The rest of the paper is organized as follows. We present the main results in Section 2 and intensive simulations in Section 3 respectively. Three real data examples are given in Section 4 and the technical proofs of the main results are given in the Appendix.

## 2 Main Results

Let $\text{SP}(\eta, \gamma)$ denote a stochastic process with mean function $\eta(t)$ and covariance function $\gamma(s,t)$. Let $y_{i1}(t), y_{i2}(t), \cdots, y_{in_i}(t)$, $i = 1, 2, \cdots, k$ be $k$ independent functional samples over a given finite time interval $\mathcal{T} = [a, b]$, $-\infty < a < b < \infty$, which satisfy

$$y_{ij}(t) = \eta_i(t) + v_{ij}(t), \; j = 1, 2, \cdots, n_i, \tag{2.1}$$
$$v_{i1}(t), v_{i2}(t), \cdots, v_{in_i}(t) \overset{i.i.d.}{\sim} \text{SP}(0, \gamma_i); \; i = 1, 2, \cdots, k,$$

where $\eta_1(t), \eta_2(t), \cdots, \eta_k(t)$ model the unknown group mean functions of the $k$ samples, $v_{ij}(t)$, $j = 1, 2, \cdots, n_i$, $i = 1, 2, \cdots, k$ denote the subject-effect functions, and $\gamma_i(s,t)$, $i = 1, 2, \cdots, k$ are the associated covariance functions. Throughout this paper, we assume that $\text{tr}(\gamma_i) < \infty$ and $\eta_i(t) \in \mathcal{L}^2(\mathcal{T})$, $i = 1, 2, \cdots, k$, where $\mathcal{L}^2(\mathcal{T})$ denotes the Hilbert space formed by all the squared integrable functions over $\mathcal{T}$ with the inner-product defined as $<f, g> = \int_{\mathcal{T}} f(t)g(t)dt$, $f, g \in \mathcal{L}^2(\mathcal{T})$. It is often of interest to test the equality of the $k$ covariance functions:

$$H_0 : \gamma_1(s,t) \equiv \gamma_2(s,t) \equiv \cdots \equiv \gamma_k(s,t), \text{ for all } s, t \in \mathcal{T}. \tag{2.2}$$

For convenience, we refer the above problem as the $k$-sample equal-covariance function (ECF) testing problem for functional data.

Based on the given $k$ functional samples (2.1), the group mean functions $\eta_i(t)$, $i = 1, 2, \cdots, k$ and the covariance functions $\gamma_i(s,t)$, $i = 1, 2, \cdots, k$ can be unbiasedly estimated as

$$\begin{aligned} \hat{\eta}_i(t) &= \bar{y}_i(t) = n_i^{-1} \sum_{j=1}^{n_i} y_{ij}(t), \; i = 1, 2, \cdots, k, \\ \hat{\gamma}_i(s,t) &= (n_i - 1)^{-1} \sum_{j=1}^{n_i} [y_{ij}(s) - \bar{y}_i(s)][y_{ij}(t) - \bar{y}_i(t)], \; i = 1, 2, \cdots, k. \end{aligned} \tag{2.3}$$

It is easy to show that $\hat{\gamma}_i(s,t)$, $i = 1, 2, \cdots, k$ are independent and $\mathrm{E}\hat{\gamma}_i(s,t) = \gamma_i(s,t)$, $i = 1, 2, \cdots, k$. Further, the estimated subject-effect functions can be written as

$$\hat{v}_{ij}(t) = y_{ij}(t) - \bar{y}_i(t), \ j = 1, 2, \cdots, n_i; \ i = 1, 2, \cdots, k. \tag{2.4}$$

When the null hypothesis (2.2) holds, let $\gamma(s,t)$ denote the common covariance function of the $k$ samples. It can be estimated by the following pooled sample covariance function

$$\hat{\gamma}(s,t) = \sum_{i=1}^{k}(n_i - 1)\hat{\gamma}_i(s,t)/(n-k), \tag{2.5}$$

where $\hat{\gamma}_i(s,t)$, $i = 1, 2, \cdots, k$ are given in (2.3) and $n = \sum_{i=1}^{k} n_i$ denotes the total sample size.

The new test we shall propose is based on the following pointwise sum of squares between groups with respect to the $k$-sample ECF testing problem (2.2):

$$SSB_n(s,t) = \sum_{i=1}^{k}(n_i - 1)[\hat{\gamma}_i(s,t) - \hat{\gamma}(s,t)]^2, \tag{2.6}$$

where $\hat{\gamma}(s,t)$ is the pooled sample covariance function of the $k$ functional samples as defined in (2.5). For each given $s,t \in \mathcal{T}$, $SSB_n(s,t)$ measures the variations of the sample covariance functions $\hat{\gamma}_i(s,t), i = 1, 2, \cdots, k$ and can be used to test the null hypothesis (2.2) restricted at $s,t \in \mathcal{T}$. Then to test the whole null hypothesis (2.2), we can use the supremum-norm of $SSB_n(s,t)$ as our test statistic:

$$T_{\max} = \sup_{s,t \in \mathcal{T}} SSB_n(s,t). \tag{2.7}$$

It is expected that when the null hypothesis is valid, $T_{\max}$ will be small and otherwise large.

To derive the asymptotic random expression of $T_{\max}$, we impose the following assumptions:

### Assumption A

1. The $k$ functional samples (2.1) are Gaussian.

2. As $n \to \infty$, the $k$ sample sizes satisfy $n_i/n \to \tau_i \in (0,1)$, $i = 1, 2, \cdots, k$.

3. The variance functions are uniformly bounded. That is, $\rho_i = \sup_{t \in \mathcal{T}} \gamma_i(t,t) < \infty$, $i = 1, 2, \cdots, k$.

The above assumptions are regular. Assumption A2 requires that the $k$ sample sizes tend to $\infty$ proportionally.

Before we state the main results, we give an alternative expression of $SSB_n(s,t)$ which is helpful for deriving the main results about $T_{\max}$. For any $s,t \in \mathcal{T}$, $SSB_n(s,t)$ can be expressed as

$$SSB_n(s,t) = \mathbf{z}_n(s,t)^T[\mathbf{I}_k - \mathbf{b}_n\mathbf{b}_n^T/(n-k)]\mathbf{z}_n(s,t), \qquad (2.8)$$

where

$$\mathbf{z}_n(s,t) = [z_1(s,t), z_2(s,t), \cdots, z_k(s,t)]^T,$$

with

$$\begin{aligned} z_i[s,t] &= \sqrt{n_i - 1}[\hat{\gamma}_i(s,t) - \gamma(s,t)], \ i=1,2,\cdots,k, \\ \mathbf{b}_n &= [\sqrt{n_1 - 1}, \sqrt{n_2 - 1}, \cdots, \sqrt{n_k - 1}]^T. \end{aligned}$$

Since $\mathbf{b}_n^T\mathbf{b}_n/(n-k) = 1$, it is easy to verify that $\mathbf{I}_k - \mathbf{b}_n\mathbf{b}_n^T/(n-k)$ is an idempotent matrix with rank $k-1$. In addition, as $n \to \infty$, we have

$$\mathbf{I}_k - \mathbf{b}_n\mathbf{b}_n^T/(n-k) \to \mathbf{I}_k - \mathbf{b}\mathbf{b}^T, \text{with } \mathbf{b} = [\sqrt{\tau_1}, \sqrt{\tau_2}, \cdots, \sqrt{\tau_k}]^T, \qquad (2.9)$$

where $\tau_i, i=1,2,\cdots,k$ are given in Assumption A2. Note that $\mathbf{I}_k - \mathbf{b}\mathbf{b}^T$ in (2.9) is also an idempotent matrix of rank $k-1$, which has the following singular value decomposition:

$$\mathbf{I}_k - \mathbf{b}\mathbf{b}^T = \mathbf{U} \begin{pmatrix} \mathbf{I}_{k-1} & \mathbf{0} \\ \mathbf{0}^T & 0 \end{pmatrix} \mathbf{U}^T, \qquad (2.10)$$

where the columns of $\mathbf{U}$ are the eigenvectors of $\mathbf{I}_k - \mathbf{b}\mathbf{b}^T$.

For further study, let $\varpi_i[(s_1,t_1),(s_2,t_2)]$ denote the covariance function between $v_{i1}(s_1)v_{i1}(t_1)$ and $v_{i1}(s_2)v_{i1}(t_2)$, $i=1,2,\cdots,k$. Then we have

$$\varpi_i[(s_1,t_1),(s_2,t_2)] = \mathrm{E}\{v_{i1}(s_1)v_{i1}(t_1)v_{i1}(s_2)v_{i1}(t_2)\} - \gamma_i(s_1,t_1)\gamma_i(s_2,t_2). \qquad (2.11)$$

Under the Gaussian assumption A1, we have

$$\varpi_i[(s_1,t_1),(s_2,t_2)] = \gamma_i(s_1,s_2)\gamma_i(t_1,t_2) + \gamma_i(s_1,t_2)\gamma_i(s_2,t_1), \ i=1,2,\cdots,k. \qquad (2.12)$$

When the null hypothesis (2.2) holds, we have

$$\varpi_i[(s_1,t_1),(s_2,t_2)] = \gamma(s_1,s_2)\gamma(t_1,t_2) + \gamma(s_1,t_2)\gamma(s_2,t_1) \equiv \varpi[(s_1,t_1),(s_2,t_2)], \ i=1,2,\cdots,k. \qquad (2.13)$$

where $\gamma(s,t)$ is the common covariance function of the $k$ functional samples. Under the above assumptions, a natural estimator of $\varpi[(s_1,t_1),(s_2,t_2)]$ is given by

$$\hat{\varpi}[(s_1,t_1),(s_2,t_2)] = \hat{\gamma}(s_1,s_2)\hat{\gamma}(t_1,t_2) + \hat{\gamma}(s_1,t_2)\hat{\gamma}(s_2,t_1). \tag{2.14}$$

Throughout this paper, let "$\xrightarrow{d}$" denote "converge in distribution" and "$X \stackrel{d}{=} Y$" denote "X and Y have the same distribution". Let $GP(\eta,\gamma)$ denote a Gaussian process with mean function $\eta$ and covariance function $\gamma$.

**Theorem 1.** *Under Assumptions A1, A2 and the null hypothesis (2.2), as $n \to \infty$, we have $T_{\max} \xrightarrow{d} T_0$ with*

$$T_0 \stackrel{d}{=} \sup_{s,t \in \mathcal{T}} \left\{ \sum_{i=1}^{k-1} w_i^2(s,t) \right\}, \tag{2.15}$$

*where $w_1(s,t), w_2(s,t), \cdots, w_{k-1}(s,t) \stackrel{i.i.d.}{\sim} GP(0,\varpi)$ with $\varpi[(s_1,t_1),(s_2,t_2)]$ defined in (2.13).*

Theorem 1 motivates us to apply a parametric bootstrap (PB) method to approximate the critical value of $T_{\max}$. This PB method can be described as follows. We firstly generate $w_i^j(s,t), \ i=1,2,\cdots,k-1; j=1,2,\cdots,N$ i.i.d. from $GP(0,\hat{\varpi})$ where $\hat{\varpi}$ is given in (2.14) and $N$ is a pre-specified large number. Then we compute $T_0^{(j)} = \sup_{s,t \in \mathcal{T}} \sum_{i=1}^{k-1}[w_i^j(s,t)]^2, j = 1,2,\cdots,N$ based on the expression (2.15). Finally, for any given significance level $\alpha$, we compute the upper $100\alpha$ sample percentile of $T_0^{(j)}, \ j=1,2,\cdots,N$ and use it as the approximate critical value of $T_{\max}$. Since the PB method makes use of the expression (2.15), it works well only when the group sample sizes $n_1, n_2, \cdots, n_k$ are large and when the functional data are Gaussian. In addition, the PB method may be time-consuming since we have to generate samples from the Gaussian process $GP(0,\hat{\varpi})$ a large number of times. This process usually takes a great deal of time as $\hat{\varpi}[(s_1,t_1),(s_2,t_2)]$ is a function on $\mathcal{T}^4$. Actually, we did conduct some preliminary simulations with the PB method. Unfortunately, we found that the above PB method is too computationally intensive even for some small sample sizes so that we have to give up our original plan to include it in our simulation studies presented in Section 3.

To overcome this difficulty, we propose a non-parametric bootstrap (NPB) method here. The key idea of the NPB method is to approximate the critical value of $T_{\max}$ via generating the bootstrapped samples from the estimated subject-effect functions (2.4). Suppose $\hat{v}_{ij}^*(t), \ j = 1,2,\cdots,n_i; \ i=1,2,\cdots,k$ are the bootstrapped $k$ samples generated from the estimated subject-effect functions (2.4). That is, each $\hat{v}_{ij}^*(t)$ takes any estimated subject-effects function from (2.4) equally likely. The NPB supremum-norm based test statistic can then be computed as

$$T^*_{\max} = sup_{s,t\in\mathcal{T}} SSB^*_n(s,t),$$

where $SSB^*_n(s,t) = \Sigma_{i=1}^k (n_i-1)[\hat{\gamma}^*_i(s,t) - \hat{\gamma}^*(s,t)]^2$, $\hat{\gamma}^*(s,t) = \sum_{i=1}^k (n_i-1)\hat{\gamma}^*_i(s,t)/(n-k)$, and $\hat{\gamma}^*_i(s,t) = (n_i-1)^{-1} \sum_{j=1}^{n_i} \hat{v}^*_{ij}(s)\hat{v}^*_{ij}(t), i=1,2,\cdots,k$. Repeating the above NPB process a large number of times, the sample upper $100\alpha$-percentile, $C^*_\alpha$, of $T^*_{\max}$ can then be computed and used as the approximate upper $100\alpha$-percentile of $T_{\max}$. The supremum-norm based test can then be conducted accordingly.

Compared with the PB method, there are a few advantages for using the NPB method. Firstly, the NPB method can be used for both small and large sample sizes. In addition, this NPB method is also applicable even though the data are not from Gaussian process. This is because the Gaussian assumption is only used in Theorem 1 to derive the asymptotic random expression of $T_{\max}$. Last but not the least, the computation of the NPB method is simple and thus may save a lot of time compared with the PB method.

**Theorem 2.** *Under Assumptions A1∼A3 and the null hypothesis (2.2), as $n \to \infty$, we have $T^*_{\max} \xrightarrow{d} T_0$ and $C^*_\alpha \xrightarrow{d} C_\alpha$ where $C_\alpha$ is the theoretical upper $100\alpha$-percentile of $T_0$.*

Theorem 2 shows that for large samples, the NPB test statistic $T^*_{\max}$ will converge in distribution to the limit random expression $T_0$ of $T_{\max}$ under the null hypothesis and hence $C^*_\alpha$ will also tend to $C_\alpha$ in distribution as $n \to \infty$. Thus it is consistent to use the NPB critical value $C^*_\alpha$ to conduct the $T_{\max}$-test.

We now study the asymptotic power of $T_{\max}$, aiming to show that the $T_{\max}$-test is root-$n$ consistent. For this end, we specify the following local alternative:

$$H_1: \gamma_i(s,t) = \gamma(s,t) + (n_i-1)^{-1/2} d_i(s,t), \ i=1,2,\cdots,k, \qquad (2.16)$$

where $d_1(s,t), d_2(s,t), \cdots, d_k(s,t)$ are some fixed bivariate functions, independent of $n$, and $\gamma(s,t)$ is some fixed covariance function. Let $\mathbf{d}(s,t) = [d_1(s,t), d_2(s,t), \cdots, d_k(s,t)]^T$. The asymptotic distribution of $T_{\max}$ and the root-$n$ consistency property with respect to the local alternative (2.16) are given in the following two theorems.

**Theorem 3.** *Under Assumptions A1∼A3 and the local alternative (2.16), as $n \to \infty$, we have*

$T_{\max} \xrightarrow{d} T_1$ with

$$T_1 \stackrel{d}{=} sup_{s,t \in \mathcal{T}} \sum_{i=1}^{k-1} [w_i(s,t) + \zeta_i(s,t)]^2, \qquad (2.17)$$

where $w_1(s,t), w_2(s,t), \cdots, w_{k-1}(s,t) \stackrel{i.i.d.}{\sim} GP(0, \varpi)$ as in Theorem 1 and $\zeta_i(s,t), i = 1, 2, \cdots, k-1$ are the $(k-1)$ components of $\boldsymbol{\zeta}(s,t) = (\mathbf{I}_{k-1}, \mathbf{0})\mathbf{U}^T \mathbf{d}(s,t)$ with $\mathbf{U}$ defined in (2.10).

Define $\delta_r^2 = ||\int_{\mathcal{T}} \int_{\mathcal{T}} \boldsymbol{\zeta}(s,t)\phi_r(s,t)dsdt||^2$, $r = 1, 2, \cdots, \infty$ with $\phi_r(s,t)$, $r = 1, 2, \cdots$ being the eigenfunctions of $\varpi[(s_1, t_1), (s_2, t_2)]$ and $\boldsymbol{\zeta}(s,t)$ defined in Theorem 3.

**Theorem 4.** *Under Assumptions A1~A3 and the local alternative (2.16), as $n \to \infty$ and $\max_r \delta_r^2 \to \infty$, the power of the supremum-norm based test, $P(T_{\max} \geq C_\alpha^*)$, will tend to 1 where $C_\alpha^*$ is the sample upper $100\alpha$-percentile of the NPB test statistic $T_{\max}^*$.*

Theorem 4 presents the root-$n$ consistency of $T_{\max}$. When the information of $\mathbf{d}(s,t)$ projected on the spaces spanned by the eigenfunctions tends to infinity, the asymptotic power of $T_{\max}$ will tend to 1. The proof of Theorem 4 is based on the following relationship between $T_{\max}$ and the $L^2$-norm based test statistic $T_n$ defined in Guo et al. (2016):

$$T_n = \int_{\mathcal{T}} \int_{\mathcal{T}} SSB_n(s,t) dsdt \leq (b-a)^2 T_{\max}, \qquad (2.18)$$

where $\mathcal{T} = [a,b]$. It then follows that

$$P(T_{\max} \geq C_\alpha^*) \geq P(T_n \geq (b-a)^2 C_\alpha^*). \qquad (2.19)$$

However, $(b-a)^2 C_\alpha^*$ may not be equal or smaller than the upper $100\alpha$-percentile of $T_n$. Thus, (2.19) does not guarantee that $T_{\max}$ has higher powers than $T_n$. Some simulation studies will be presented in the next section to compare the powers of $T_{\max}$ and $T_n$ under various simulation configurations.

## 3 Simulation Studies

For the ECF testing problem (2.2), Guo et al. (2016) proposed an $L^2$-norm based test whose null distribution can be approximated by a naive method, a bias-reduced method and a random permutation method. The associated $L^2$-norm based tests may be denoted as $L_{nv}^2, L_{br}^2$ and $L_{rp}^2$ respectively for easy reference. In this section, we present some simulation studies, aiming to compare the $T_{\max}$-test against $L_{nv}^2, L_{br}^2$ and $L_{rp}^2$ under various simulation configurations. In the simulation studies, we generate $k$ functional samples using the following data generating model:

$$y_{ij}(t) = \eta_i(t) + v_{ij}(t), \ \eta_i(t) = \sum_{r=1}^{q} c_{ir} t^{r-1}, \ v_{ij}(t) = \mathbf{b}_{ij}^T \boldsymbol{\Psi}_i(t), \ t \in [0,1],$$
$$\mathbf{b}_{ij} = [b_{ij1}, b_{ij2}, \cdots, b_{ijq}]^T, \ b_{ijr} \stackrel{d}{=} \sqrt{\lambda_r} z_{ijr}, \ r = 1, 2, \cdots, q, \ j = 1, 2, \cdots, n_i, \ i = 1, 2, \cdots, k,$$
(3.1)

where the parameters $c_{ir}, \ i = 1, 2 \cdots, k, \ r = 1, 2, \cdots, q$ for the group mean functions $\eta_i(t), \ i = 1, 2, \cdots, k$, can be flexibly specified, the random variables $z_{ijr}, \ r = 1, 2, \cdots, q; \ j = 1, 2, \cdots, n_i; \ i = 1, 2, \cdots, k$ are i.i.d. with mean 0 and variance 1, $\boldsymbol{\Psi}_i(t) = [\psi_{i1}(t), \psi_{i2}(t), \cdots, \psi_{iq}(t)]^T$ is a vector of $q$ basis functions $\psi_{ir}(t), \ t \in [0,1], \ r = 1, 2, \cdots, q$, and the variance components $\lambda_r, \ r = 1, 2, \cdots, q$ are positive and decreasing in $r$, and the number of the basis functions $q$ is an odd positive integer. These tuning parameters help specify the covariance functions

$$\gamma_i(s,t) = \boldsymbol{\Psi}_i(s)^T diag(\lambda_1, \lambda_2, \cdots, \lambda_q) \boldsymbol{\Psi}_i(t) = \sum_{r=1}^{q} \lambda_r \psi_{ir}(s) \psi_{ir}(t), i = 1, 2, \cdots, k.$$

We also assume that the design time points for all the functions $y_{ij}(t), \ j = 1, 2, \cdots, n_i, \ i = 1, 2, \cdots, k$ are the same and are specified as $t_j = (j-1)/(J-1), \ j = 1, 2, \cdots, J$, where $J$ is some positive integer. If the sampling time points are different across various functions or the sampling time points are not equally spaced, some smoothing techniques, e.g., local polynomial kernel smoothing (Fan and Gijbels 1996), smoothing splines (Eubank 1999) and P-splines (Ruppert et al. 2003) among others can be applied to reconstruct the individual functional observations. We shall not consider these smoothing techniques in the simulations conducted in this section for time saving.

We now specify the model parameters in (3.1). The number of groups is chosen as $k = 3$. To specify the group mean functions $\eta_1(t), \eta_2(t), \cdots, \eta_k(t)$, we set $c_{ir} = (\frac{1}{2})^{i-1} r, \ r = 1, 2, \cdots, q$ (Actually, the mean functions $\eta_i(t), i = 1, 2, \cdots, k$ can be directly specified as 0 since the tests under consideration are independent of the specification of the mean functions $\eta_i(t), i = 1, 2, \cdots, k$.) Then we specify the covariance functions $\gamma_i(s,t), i = 1, 2, \cdots, k$ in the following way. First, we set $\lambda_r = \rho^{r-1}, \ r = 1, 2, \cdots, q$ for $0 < \rho < 1$. Then, we select a vector of $q$ orthonormal Fourier basis functions, denoted as $\boldsymbol{\Phi}(t) = [\phi_1(t), \phi_2(t), \cdots, \phi_q(t)]$ where

$$\phi_1(t) = 1, \ \phi_{2r}(t) = \sqrt{2}\sin(2\pi r t), \ \phi_{(2r+1)}(t) = \sqrt{2}\cos(2\pi r t), t \in [0,1], r = 1, 2, \cdots, (q-1)/2.$$

To obtain the $k$ different basis function vectors $\boldsymbol{\Psi}_i(t), i = 1, 2, \cdots, k$, we set $\psi_{ir}(t) = \phi_r(t)$, $r = 1, 3, 4 \cdots q$, and $\psi_{i2}(t) = \phi_2(t) + (i-1)\omega, \ i = 1, 2, \cdots, k$ for simplicity. With these basis function vectors $\boldsymbol{\Psi}_i(t), i = 1, 2, \cdots, k$, we have $k$ different covariance functions

$$\gamma_i(s,t) = \gamma_1(s,t) + (i-1)\lambda_2(\phi_2(s) + \phi_2(t))\omega + (i-1)^2\lambda_2\omega^2, i = 1, 2, \cdots, k.$$

Note that the differences of the $k$ covariance functions are located in the space spanned by the first two basis functions $\phi_1(t), \phi_2(t), t \in [0,1]$ of the basis function vector $\boldsymbol{\Phi}(\mathbf{t})$ and these differences are controlled by the tuning parameter $\omega$. Notice also that the tuning parameter $\rho$ not only determines the decay rate of $\lambda_1, \lambda_2, \cdots, \lambda_q$, but also determines how the simulated functional data are correlated: when $\rho$ is close to 0, $\lambda_1, \lambda_2, \cdots, \lambda_q$ will decay very fast, indicating that the simulated functional data are highly correlated; and when $\rho$ is close to 1, $\lambda_r$, $r = 1, 2, \cdots, q$ will decay slowly, indicating that the simulated functional data are nearly uncorrelated. In addition, we set $q = 21$ and the number of design time points $J = 180$. We also set $\rho = 0.1, 0.3, 0.5$ to consider the three correlation cases when the simulated functional data have very high, high and moderate correlations and specify three cases of the sample size vector: $\mathbf{n}_1 = [n_1, n_2, n_3] = [20, 30, 30]$, $\mathbf{n}_2 = [30, 40, 50]$ and $\mathbf{n}_3 = [80, 70, 100]$, representing the small, moderate and large sample size cases respectively. We choose those three types of correlation because most functional data have high correlations. Finally, we specify two cases of the distribution of the i.i.d. random variables $z_{ijr}$, $r = 1, 2, \cdots, q$; $j = 1, 2, \cdots, n_i$; $i = 1, 2, \cdots, k$: $z_{ijr} \overset{i.i.d.}{\sim} N(0,1)$ and $z_{ijr} \overset{i.i.d.}{\sim} t_4/\sqrt{2}$, allowing to generate Gaussian and non-Gaussian functional data respectively with $z_{ijr}$ having mean 0 and variance 1. Notice that the $t_4/\sqrt{2}$ distribution is chosen since it has nearly heaviest tails among the $t$-distributions with finite first two moments.

For a given model configuration, the $k = 3$ functional samples (3.1) are generated. We then apply $L^2_{nv}, L^2_{br}, L^2_{rp}$ and $T_{\max}$ to them to test the ECF testing problem (2.2) and their p-values are computed respectively. In particular, the p-values of $L^2_{rp}$ and $T_{\max}$ are obtained via 500 runs of random permutations or nonparametric bootstrapping. We reject the null hypothesis (2.2) if the resulting p-value of a testing procedure is smaller than the nominal significance level $\alpha = 5\%$. Repeat the above simulation process, 10000 times, say, so that the associated empirical sizes or powers can be obtained.

We are now ready to check how $T_{\max}$ performs compared with $L^2_{nv}, L^2_{br}$, and $L^2_{rp}$ in terms of level accuracy and power. Table 1 displays the empirical sizes and powers (in percentages) of $L^2_{nv}, L^2_{br}, L^2_{rp}$ and $T_{\max}$ when $z_{ijr}$, $r = 1, 2, \cdots, q$; $j = 1, 2, \cdots, n_i$; $i = 1, 2, \cdots, k$ are i.i.d. $N(0,1)$. It is seen that in terms of level accuracy, $L^2_{rp}$ and $T_{\max}$ are generally comparable with their empirical sizes being slightly liberal while $L^2_{nv}$ and $L^2_{br}$ are comparable with their empirical sizes being slightly conservative. However, in terms of power, $T_{\max}$ generally has higher powers than $L^2_{nv}, L^2_{br}$, and $L^2_{rp}$ when the functional data are highly correlated ($\rho = 0.1, 0.3$). This shows that $T_{\max}$ is advantageous since functional data are generally highly correlated. Of course, it is also seen that $T_{\max}$ has lower powers than $L^2_{nv}, L^2_{br}$, and $L^2_{rp}$ when the functional data are moderately correlated ($\rho = 0.5$) but this situation may be improved with the sample sizes enlarged. Note the fact that $T_{\max}$ is less powerful compared with $L^2_{nv}, L^2_{br}$ and $L^2_{rp}$ when the functional data are less correlated is not a surprise since when the functional data are less correlated, $T_{\max}$ just uses the information at the supremum value while $L^2_{nv}, L^2_{br}$ and $L^2_{rp}$ can take more information into account via the $L^2$-norm of the differences between the individual sample covariance functions and the pooled covariance function.

Table 1: Empirical sizes and powers (in percentages) of $L^2_{nv}, L^2_{br}, L^2_{rp}$ and $T_{\max}$ when $z_{ijr}$, $r = 1, \cdots, q$; $j = 1, \cdots, n_i$; $i = 1, \cdots, k$ are i.i.d. $N(0,1)$.

|  |  | $\mathbf{n}_1 = [20, 30, 30]$ | | | | $\mathbf{n}_2 = [30, 40, 50]$ | | | | | $\mathbf{n}_3 = [80, 70, 100]$ | | | | |
|---|---|---|---|---|---|---|---|---|---|---|---|---|---|---|---|
| $\rho$ | $\omega$ | $L^2_{nv}$ | $L^2_{br}$ | $L^2_{rp}$ | $T_{max}$ | $\omega$ | $L^2_{nv}$ | $L^2_{br}$ | $L^2_{rp}$ | $T_{max}$ | $\omega$ | $L^2_{nv}$ | $L^2_{br}$ | $L^2_{rp}$ | $T_{max}$ |
|  | 0.00 | 4.89 | 5.15 | 6.01 | 5.72 | 0.00 | 4.73 | 4.95 | 5.84 | 5.70 | 0.00 | 4.51 | 4.62 | 5.27 | 4.76 |
|  | 1.00 | 10.82 | 11.33 | 12.43 | 19.45 | 1.20 | 25.90 | 26.42 | 26.41 | 43.02 | 1.00 | 38.88 | 39.24 | 39.07 | 66.82 |
| 0.1 | 2.00 | 54.41 | 55.38 | 52.40 | 59.48 | 1.60 | 51.34 | 52.00 | 51.23 | 66.49 | 1.20 | 60.31 | 60.64 | 60.08 | 83.13 |
|  | 3.00 | 89.63 | 90.12 | 86.81 | 87.28 | 2.20 | 86.16 | 86.62 | 84.18 | 90.72 | 1.50 | 87.09 | 87.25 | 86.76 | 95.89 |
|  | 6.00 | 99.98 | 99.98 | 99.69 | 99.47 | 2.80 | 97.55 | 97.65 | 96.40 | 97.82 | 2.00 | 99.29 | 99.30 | 99.22 | 99.81 |
|  | 0.00 | 4.37 | 4.74 | 5.58 | 5.79 | 0.00 | 4.25 | 4.45 | 5.05 | 5.62 | 0.00 | 4.94 | 5.03 | 5.02 | 5.17 |
|  | 0.80 | 28.53 | 29.50 | 29.26 | 31.65 | 0.60 | 25.22 | 25.92 | 26.80 | 33.60 | 0.30 | 14.28 | 14.57 | 15.27 | 24.94 |
| 0.3 | 1.20 | 63.31 | 64.44 | 61.83 | 58.38 | 0.90 | 60.08 | 61.00 | 59.93 | 64.30 | 0.50 | 44.01 | 44.44 | 44.64 | 57.34 |
|  | 1.80 | 93.18 | 93.49 | 90.30 | 85.23 | 1.20 | 86.53 | 87.00 | 85.02 | 84.69 | 0.80 | 91.73 | 91.86 | 90.61 | 93.39 |
|  | 2.50 | 99.29 | 99.37 | 98.02 | 95.31 | 1.40 | 94.90 | 95.19 | 93.77 | 92.70 | 1.00 | 98.83 | 98.84 | 98.61 | 98.84 |
|  | 0.00 | 4.32 | 4.98 | 5.91 | 5.88 | 0.00 | 4.47 | 4.85 | 5.63 | 6.03 | 0.00 | 5.04 | 5.26 | 5.31 | 5.18 |
|  | 0.50 | 22.09 | 23.14 | 24.62 | 20.11 | 0.40 | 24.30 | 25.58 | 26.32 | 22.86 | 0.30 | 35.96 | 36.63 | 36.61 | 32.66 |
| 0.5 | 0.80 | 54.47 | 56.15 | 54.03 | 40.58 | 0.60 | 53.21 | 54.47 | 54.01 | 45.27 | 0.40 | 62.09 | 62.66 | 61.82 | 53.41 |
|  | 1.00 | 73.82 | 75.30 | 71.44 | 55.70 | 0.90 | 88.42 | 88.92 | 86.66 | 75.35 | 0.45 | 74.07 | 74.52 | 73.47 | 64.11 |
|  | 2.00 | 99.61 | 99.73 | 98.56 | 93.00 | 1.20 | 98.52 | 98.59 | 97.84 | 92.31 | 0.70 | 98.49 | 98.55 | 98.12 | 94.36 |

When the functional data are non-Gaussian, similar conclusions can also be obtained except now $L^2_{nv}$ and $L^2_{br}$ are no longer workable since their empirical sizes are too large compared with the nominal size 5%. Table 2 shows the empirical sizes and powers of $L^2_{nv}, L^2_{br}, L^2_{rp}$ and $T_{\max}$ when $z_{ijr}$, $r = 1, 2, \cdots, q$; $j = 1, 2, \cdots, n_i$; $i = 1, 2, \cdots, k \overset{i.i.d.}{\sim} t_4/\sqrt{2}$. It is seen that in terms of level accuracy, $L^2_{rp}$ and $T_{\max}$ are generally comparable with their empirical sizes being slightly liberal and $L^2_{nv}$ and $L^2_{br}$ have very large empirical sizes which show that $L^2_{nv}$ and $L^2_{br}$ are not applicable for non-Gaussian functional data and hence it does not make any sense to compare their powers with $T_{\max}$ and $L^2_{rp}$. We thus just compare the empirical powers of $T_{\max}$ with $L^2_{rp}$. We see that $T_{\max}$ generally has higher powers than $L^2_{rp}$ when the functional data are highly correlated ($\rho = 0.1, 0.3$) except when $\boldsymbol{n}_1 = [20, 30, 30]$ which may be too small for $T_{\max}$ to work properly. We also see that $T_{\max}$ has lower powers than $L^2_{rp}$ when the functional data are moderately correlated ($\rho = 0.5$).

Table 2: Empirical sizes and powers (in percentages) of $L^2_{nv}, L^2_{br}, L^2_{rp}$ and $T_{\max}$ when $z_{ijr}$, $r = 1, \cdots, q$; $j = 1, \cdots, n_i$; $i = 1, \cdots, k$ are i.i.d. $t_4/\sqrt{2}$.

| | | $\mathbf{n}_1 = [20, 30, 30]$ | | | | $\mathbf{n}_2 = [30, 40, 50]$ | | | | $\mathbf{n}_3 = [80, 70, 100]$ | | | |
|---|---|---|---|---|---|---|---|---|---|---|---|---|---|
| $\rho$ | $\omega$ | $L^2_{nv}$ | $L^2_{br}$ | $L^2_{rp}$ | $T_{max}$ | $\omega$ | $L^2_{nv}$ | $L^2_{br}$ | $L^2_{rp}$ | $T_{max}$ | $\omega$ | $L^2_{nv}$ | $L^2_{br}$ | $L^2_{rp}$ | $T_{max}$ |
| | 0.00 | 30.50 | 31.23 | 6.31 | 5.82 | 0.00 | 33.47 | 34.02 | 6.35 | 6.14 | 0.00 | 41.15 | 41.28 | 5.15 | 5.22 |
| | 1.50 | 45.44 | 46.21 | 18.70 | 22.92 | 1.50 | 58.48 | 59.07 | 28.22 | 36.86 | 1.00 | 62.29 | 62.59 | 20.44 | 36.42 |
| 0.1 | 3.00 | 84.66 | 85.36 | 57.72 | 55.64 | 2.20 | 82.28 | 82.76 | 56.29 | 62.08 | 1.50 | 85.74 | 85.94 | 52.42 | 67.75 |
| | 6.00 | 99.26 | 99.29 | 83.58 | 80.39 | 3.50 | 97.87 | 97.95 | 86.44 | 87.03 | 2.00 | 96.43 | 96.46 | 78.54 | 86.16 |
| | 12.00 | 100.00 | 100.00 | 89.29 | 87.43 | 6.00 | 99.88 | 99.88 | 95.84 | 95.22 | 3.00 | 99.84 | 99.84 | 96.00 | 96.53 |
| | 0.00 | 33.52 | 34.53 | 6.19 | 6.03 | 0.00 | 37.56 | 38.20 | 6.30 | 5.95 | 0.00 | 45.47 | 45.68 | 5.91 | 5.21 |
| | 1.00 | 60.70 | 61.81 | 28.50 | 25.66 | 0.50 | 48.14 | 49.09 | 13.25 | 16.40 | 0.50 | 72.45 | 72.62 | 22.50 | 31.14 |
| 0.3 | 2.00 | 91.76 | 92.15 | 66.33 | 56.61 | 1.20 | 85.14 | 85.55 | 57.35 | 53.97 | 0.80 | 91.25 | 91.40 | 55.74 | 61.17 |
| | 5.00 | 99.91 | 99.91 | 88.69 | 83.70 | 2.00 | 98.24 | 98.29 | 86.93 | 82.75 | 1.00 | 97.11 | 97.17 | 75.11 | 76.50 |
| | 8.00 | 99.99 | 99.99 | 89.59 | 86.79 | 5.00 | 99.95 | 99.95 | 97.13 | 96.15 | 2.00 | 99.98 | 99.98 | 97.99 | 97.46 |
| | 0.00 | 39.83 | 41.47 | 7.06 | 6.14 | 0.00 | 43.93 | 45.08 | 6.67 | 6.13 | 0.00 | 54.27 | 54.74 | 5.45 | 5.31 |
| | 0.80 | 67.55 | 68.89 | 31.99 | 22.32 | 0.50 | 66.31 | 67.40 | 24.04 | 19.49 | 0.40 | 85.01 | 85.31 | 29.41 | 27.38 |
| 0.5 | 1.50 | 93.38 | 93.79 | 66.62 | 49.29 | 0.90 | 88.82 | 89.27 | 58.32 | 44.77 | 0.60 | 95.65 | 95.79 | 60.72 | 51.70 |
| | 2.00 | 97.89 | 98.03 | 77.52 | 61.53 | 1.70 | 99.22 | 99.25 | 89.85 | 80.75 | 1.00 | 99.84 | 99.84 | 91.13 | 85.05 |
| | 6.00 | 99.99 | 99.99 | 90.14 | 85.66 | 4.00 | 99.98 | 99.98 | 96.94 | 95.23 | 1.50 | 100.00 | 100.00 | 97.61 | 95.81 |

In some situations, $T_{\max}$ can have much higher powers than $L^2_{nv}, L^2_{br}$, and $L^2_{rp}$ even when functional data are moderately correlated. To show this is the case, we just need to make a small change of the simulation settings used earlier. We continue to use the data generating model (3.1) and set $\eta_i(t) = 0, i = 1, 2, \cdots, k$. However, we now set $\psi_{i1}(t) = \phi_1(t) + (i-1)\frac{2}{\sqrt{\pi}}e^{-4t^2}\omega, i = 1, 2, \cdots, k$ so that the differences of the basis function vectors $\Psi_i(t), i = 1, 2, \cdots, k$ are now located at the first basis function. Under this new scheme, we conduct a simulation study which is similar to the one which yielded Table 1. Table 3 displays the empirical sizes and powers (in percentages) of $L_{nv}^2, L_{br}^2, L_{rp}^2$ and $T_{\max}$ when $z_{ijr}$, $r = 1, 2, \cdots, q$; $j = 1, 2, \cdots, n_i$; $i = 1, 2, \cdots, k$ are i.i.d. $N(0,1)$ under the new simulation scheme. It is seen that $T_{max}$ has much higher powers than $L_{nv}^2, L_{br}^2$ and $L_{rp}^2$ for $\rho = 0.1, 0.3$, and $0.5$ as well.

Table 3: Empirical sizes and powers (in percentages) of $L_{nv}^2, L_{br}^2, L_{rp}^2$ and $T_{\max}$ when $z_{ijr}$, $r = 1, \cdots, q$; $j = 1, \cdots, n_i$; $i = 1, \cdots, k$ are i.i.d. $N(0,1)$ under the new simulation scheme.

| | | $\mathbf{n}_1$=[20,30,30] | | | | $\mathbf{n}_2$=[30,40,50] | | | | $\mathbf{n}_3$=[80,70,100] | | | |
|---|---|---|---|---|---|---|---|---|---|---|---|---|---|
| $\rho$ | $\omega$ | $L_{nv}^2$ | $L_{br}^2$ | $L_{rp}^2$ | $T_{max}$ | $\omega$ | $L_{nv}^2$ | $L_{br}^2$ | $L_{rp}^2$ | $T_{max}$ | $\omega$ | $L_{nv}^2$ | $L_{br}^2$ | $L_{rp}^2$ | $T_{max}$ |
| | 0.00 | 4.04 | 4.31 | 5.49 | 5.49 | 0.00 | 4.85 | 4.93 | 5.63 | 5.73 | 0.00 | 4.46 | 4.54 | 5.13 | 5.22 |
| | 0.07 | 5.03 | 5.31 | 6.31 | 23.59 | 0.05 | 5.30 | 5.45 | 6.24 | 21.36 | 0.03 | 5.67 | 5.79 | 5.93 | 19.91 |
| 0.1 | 0.15 | 10.24 | 10.83 | 11.92 | 64.79 | 0.10 | 8.64 | 8.90 | 9.53 | 62.05 | 0.05 | 7.80 | 7.91 | 8.17 | 54.40 |
| | 0.21 | 18.86 | 19.52 | 21.21 | 84.04 | 0.15 | 16.95 | 17.46 | 18.62 | 88.55 | 0.07 | 10.87 | 11.05 | 11.56 | 82.59 |
| | 0.42 | 72.04 | 73.25 | 70.95 | 98.86 | 0.20 | 30.94 | 31.60 | 32.66 | 97.35 | 0.10 | 19.90 | 20.24 | 20.99 | 97.95 |
| | 0.00 | 4.25 | 4.53 | 5.45 | 5.30 | 0.00 | 4.22 | 4.42 | 5.10 | 5.57 | 0.00 | 5.37 | 5.54 | 5.77 | 5.51 |
| | 0.09 | 6.36 | 6.76 | 7.80 | 23.68 | 0.07 | 5.41 | 5.83 | 6.44 | 21.20 | 0.05 | 7.92 | 8.12 | 8.68 | 28.11 |
| 0.3 | 0.17 | 13.15 | 13.99 | 15.37 | 63.99 | 0.11 | 9.13 | 9.44 | 10.60 | 55.02 | 0.07 | 11.71 | 11.89 | 12.25 | 59.84 |
| | 0.24 | 25.44 | 26.73 | 28.20 | 84.14 | 0.18 | 24.05 | 24.83 | 25.10 | 91.33 | 0.09 | 16.78 | 17.03 | 17.34 | 85.16 |
| | 0.44 | 73.51 | 74.96 | 72.60 | 98.35 | 0.21 | 33.17 | 34.12 | 34.36 | 95.66 | 0.11 | 25.92 | 26.44 | 26.27 | 95.91 |
| | 0.00 | 4.31 | 4.74 | 6.03 | 5.66 | 0.00 | 4.57 | 5.01 | 5.60 | 5.21 | 0.00 | 4.51 | 4.66 | 5.00 | 5.06 |
| | 0.10 | 6.04 | 6.84 | 8.21 | 14.91 | 0.10 | 8.52 | 9.05 | 9.31 | 24.58 | 0.05 | 8.01 | 8.22 | 8.29 | 12.47 |
| 0.5 | 0.20 | 16.12 | 17.38 | 18.98 | 56.78 | 0.15 | 15.45 | 16.53 | 17.31 | 57.44 | 0.08 | 13.37 | 13.81 | 13.77 | 40.84 |
| | 0.30 | 36.68 | 38.47 | 38.49 | 84.01 | 0.22 | 33.95 | 35.13 | 35.32 | 87.98 | 0.10 | 20.17 | 20.82 | 20.76 | 66.48 |
| | 0.50 | 81.75 | 83.20 | 79.06 | 97.92 | 0.29 | 60.08 | 61.41 | 59.54 | 97.68 | 0.15 | 47.73 | 48.49 | 47.81 | 96.81 |

In the above three simulation studies, we see that $T_{\max}$ generally has higher powers than $L_{nv}^2, L_{br}^2$ and $L_{rp}^2$ when the functional data have higher or even moderate correlation and when the sample sizes are large enough, and it has lower powers when the functional data have lower correlation or when the sample sizes are too small.

## 4 Applications to Three Real Data Examples

In this section, we shall present the application of $T_{\max}$, together with $L_{nv}^2, L_{br}^2, L_{rp}^2$, to three real data examples. From these three examples, we shall see that $T_{\max}$ often has higher power than $L_{nv}^2, L_{br}^2, L_{rp}^2$ in detecting the covariance function differences of different functional populations.

## 4.1 Canadian Temperature Data

Figure 1: *Reconstructed individual temperature functions of the Canadian temperature data.*

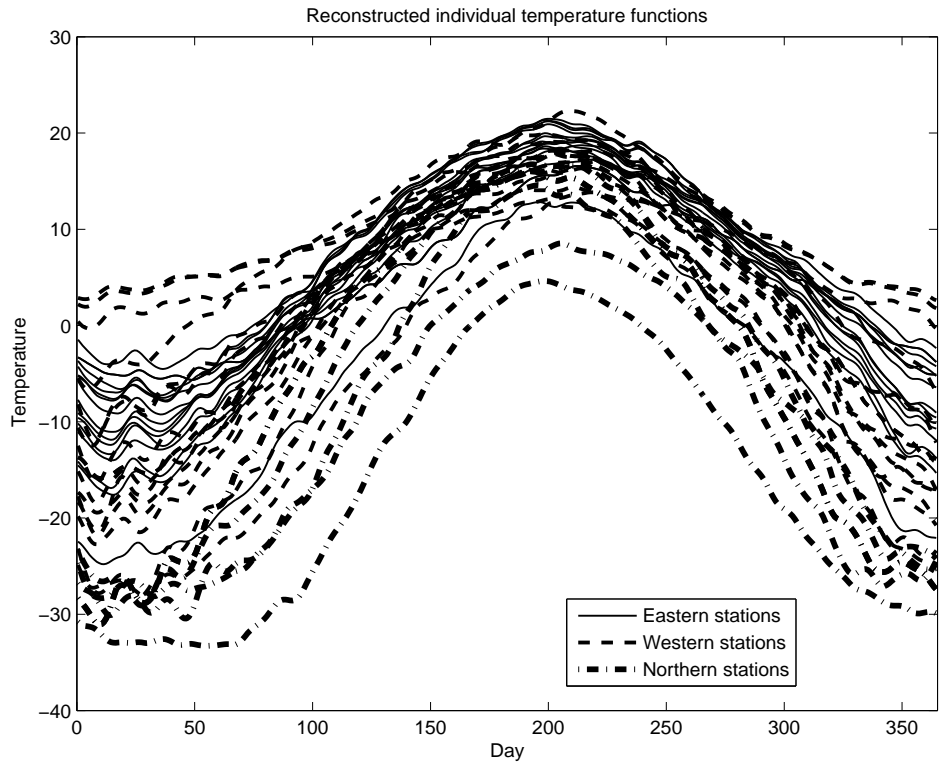

The Canadian temperature data set has been used for illustrating various methodologies for functional data; see for example, Ramsay and Silverman (2005), Zhang and Chen (2007) and Zhang and Liang (2013) among others. The temperature functional observations consist of daily temperature records of 35 weather stations over 365 days, with each observation being a temperature curve as shown in Figure 1 where the reconstructed individual temperature functions over a whole year are depicted. These weather stations are located in three different regions over Canada. There are 15 weather stations located in eastern Canada, another 15 in western Canada and the remaining 5 in northern Canada. We are interested in the equality of the covariance functions (variations) of the temperature functions at the three different regions over the whole year and four seasons (spring (March, April, and May or $J = [60, 151]$), summer (June, July, and August or $J = [152, 243]$), autumn (September, October and November or $J = [244, 334]$) and winter (December, January and February or $J = [1, 59] \cup [335, 365]$).

Table 4: P-values of $L_{nv}^2, L_{br}^2, L_{rp}^2$ and $T_{\max}$ for testing the equality of the covariance functions for the Canadian temperature data.

| $\mathcal{T}$ | $L_{nv}^2$ | $L_{br}^2$ | $L_{rp}^2$ | $T_{max}$ |
|---|---|---|---|---|
| Whole | 0.0383 | 0.0323 | 0.0451 | 0.0321 |
| Spring | 0.0224 | 0.0193 | 0.0621 | 0.1228 |
| Summer | 0.0010 | 0.0009 | 0.0019 | 0.0199 |
| Autumn | 0.1997 | 0.1816 | 0.1896 | 0.0917 |
| Winter | 0.0266 | 0.0234 | 0.0297 | 0.0339 |

Table 4 shows the p-values of $L_{nv}^2, L_{br}^2, L_{rp}^2$ and $T_{\max}$ for testing the equality of the covariance functions (variations) of the Canadian temperature functions of the eastern, western and northern weather stations over the whole year and the four seasons (spring, summer, autumn, and winter). The p-values of $L_{rp}^2$ and $T_{\max}$ were obtained via 10000 runs of random permutations and nonparametric bootstrapping respectively. It can be seen that all the tests suggest that the covariance functions of the three regions over the whole year and in summer and winter are unlikely to be the same but they may be quite similar in autumn. The testing results in spring are not consistent. $L_{nv}^2$ and $L_{br}^2$ suggest that the covariance functions of the three regions in spring is unlikely to be the same but $L_{rp}^2$ and $T_{\max}$ are not so sure. Since $L_{nv}^2$ and $L_{br}^2$ only work under the assumption that the functional data are Gaussian while $L_{rp}^2$ and $T_{\max}$ do not make such an assumption, the testing results of $L_{rp}^2$ and $T_{\max}$ are more reliable than those of $L_{nv}^2$ and $L_{br}^2$.

### 4.2 Nitrogen Oxide Emission Level Data

We now present the application of $T_{\max}$, together with $L_{nv}^2, L_{br}^2, L_{rp}^2$, to another data set consisting of Nitrogen oxides (NOx) emission levels (in $\mu g/m^3$) measured by an environmental control station close to an industrial area in Poblenou, Barcelona, Spain. The NOx emission level data were kindly made available by Febrero et al. (2008). Each curve of the NOx level data was recorded every hour per day from February 23 to June 26 in 2005. The data set has been studied in Febrero et al. (2008) for illustrating an outlier detection method. In large cities, especially those with heavy traffic and well-developed industries, NOx gases are known to be among the most important pollutants and thus the emission levels of nitrogen oxides (NOx) can be significant. The NOx emission level curves of the data set may be classified into two groups according to the working days and non-working days. The working day group includes 76 NOx emission level curves while the non-working day group has 39 curves. Since the NOx gases are mainly emitted into the atmosphere in sources of motor vehicles and industries, we are wondering if the covariance functions of the working day group and the non-working day group are the same.

Figure 2: *3-D plots of the estimated covariance functions of the NOx emission level data.*

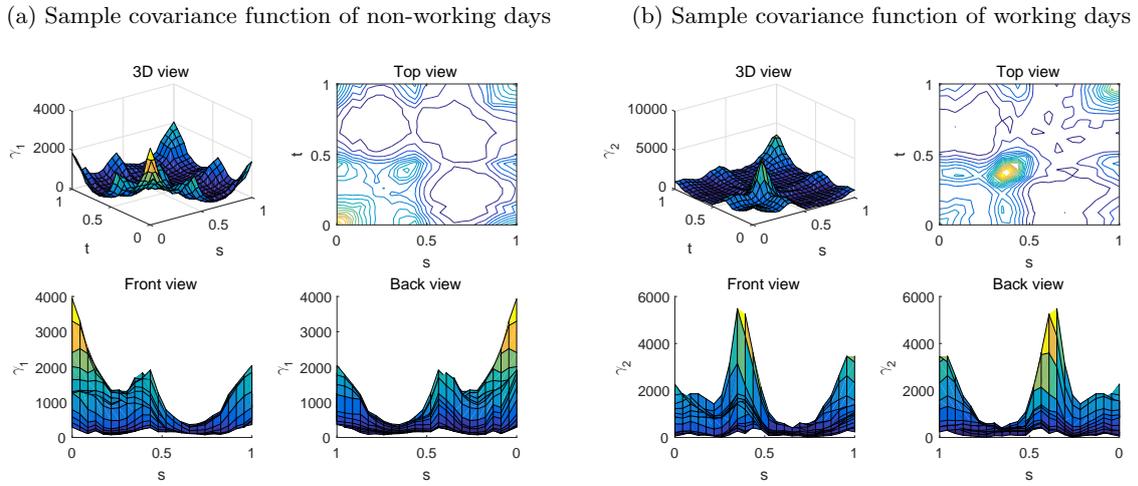

(a) Sample covariance function of non-working days    (b) Sample covariance function of working days

Figure 2 shows the 3-D plots of the estimated covariance functions of the NOx emission level curves of working days and non-working days. It seems that the two sample covariance functions are not the same. We then applied $T_{\max}$, together with $L_{nv}^2, L_{br}^2$ and $L_{rp}^2$, to check if the differences of the covariance functions of the NOx emission level curves of working days and non-working days are significant. The p-values of $T_{\max}$ is $0.006$ while those of $L_{nv}^2, L_{br}^2$ and $L_{rp}^2$ are $0.1193, 0.1133$ and $0.3427$ respectively. The p-values of $T_{\max}$ and $L_{rp}^2$ were obtained via 10000 runs of nonparametric bootstrapping and random permutations respectively. It is seen that $T_{\max}$ can detect the differences of the covariance functions between the NOx emission level curves of working days and non-working days. This is consistent with what we observed from Figure 2. However, $L_{nv}^2, L_{br}^2$ and $L_{rp}^2$ cannot. This shows that $T_{\max}$ is indeed more powerful than $L_{nv}^2, L_{br}^2$ and $L_{rp}^2$ in detecting the covariance function differences between the working day and non-working day groups.

## 4.3 Berkeley Growth Data

We finally present the application of $T_{\max}$, together with $L_{nv}^2, L_{br}^2, L_{rp}^2$, to the Berkeley growth curve data set which has been extensively studied in Ramsay and Silverman (2005) and Ramsay and Silverman (2002). This data set contains the heights of 39 boys and 54 girls from age 1 to 18 (Tuddenham and Snyder 1954). Notice that the 31 ages at which the data were collected are not equally spaced. It is of interest to check whether the variable "gender" has some impact on the covariance structure of a child's grow curve. In other words, we want to test the equality of the covariance functions of boys' and girls' growth curves.

Figure 3: *3-D plots of the estimated covariance functions of the Berkeley growth curve data.*

(a) Sample covariance function of the heights of boys    (b) Sample covariance function of the heights of girls

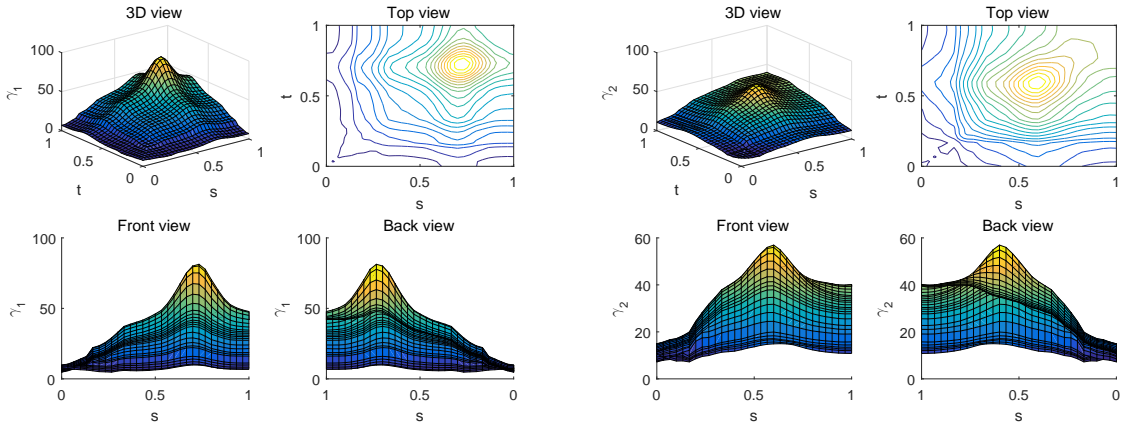

Figure 3 depicts the 3-D plots of the estimated covariance functions of the Berkeley growth curve data. It seems that there is a clear difference between the sample covariance structures of boys and girls. To verify if this is the case, we applied $T_{\max}$, together with $L_{nv}^2, L_{br}^2$ and $L_{rp}^2$. The p-values of $T_{\max}$ is 0.0453 while those of $L_{nv}^2, L_{br}^2$ and $L_{rp}^2$ are $0.4649, 0.4571$ and $0.4762$ respectively. Again, the p-values of $T_{\max}$ and $L_{rp}^2$ were obtained via 10000 runs of nonparametric bootstrapping and random permutations respectively. It is seen that $T_{\max}$ can detect the differences of the covariance functions of the growth curves of boys and girls. This is consistent with what we observed from Figure 3. However, $L_{nv}^2, L_{br}^2$ and $L_{rp}^2$ cannot. This shows that $T_{\max}$ is again more powerful than $L_{nv}^2, L_{br}^2$ and $L_{rp}^2$ in detecting the covariance function differences between the growth curves of boys and girls.

# Appendix

Technical proofs and additional contents are available in supplementary materials.